\documentclass[pra,reprint,superscriptaddress]{revtex4-1}

\usepackage{amsmath,amsfonts,amssymb,amsthm,braket,graphicx,color,lipsum}
\usepackage[colorlinks=true,citecolor=blue,linkcolor=magenta]{hyperref}
\usepackage{bm}

\usepackage{aligned-overset}

\usepackage[normalem]{ulem}
\usepackage[dvipsnames]{xcolor}

\newcommand{\todo}[1] {}

\global\long\def\avg#1{\langle#1\rangle}
\global\long\def\p{\prime}

\global\long\def\ket#1{|#1\rangle}
\global\long\def\bra#1{\langle#1|}
\global\long\def\proj#1#2{|#1\rangle\langle#2|}
\global\long\def\inner#1#2{\langle#1|#2\rangle}
\global\long\def\tr{\mathrm{tr}}

\global\long\def\im{\imath}

\newcommand{\dg} {{\dagger}}
\newcommand{\pd} {{\phantom\dagger}}

\newcommand{\ci}[1] {{c_{#1}^{\pd}}}
\newcommand{\cid}[1] {c_{#1}^\dg}

\newcommand{\bG} {\bm{G}}
\newcommand{\bS} {\bm{\Sigma}}
\newcommand{\bGa} {\bm{\Gamma}}
\newcommand{\bH}{\bm{\bar{H}}}
\newcommand{\bHS}{\bm{\bar{H}}_\qs}

\renewcommand{\Im} {\operatorname{Im}}

\newcommand{\ql}{\mathcal{L}}
\newcommand{\qs}{\mathcal{S}}
\newcommand{\qr}{\mathcal{R}}

\newcommand{\qc}{\mathcal{C}}

\newcommand{\ok}{\proj{v_k}{v_k}}
\newcommand{\bk}{\bra{v_k}}
\newcommand{\kk}{\ket{v_k}}
\newcommand{\ik}{\left| v_k \right|^2} 

\newcommand{\bkFoot}{\protect\langle v_k|}
\newcommand{\kkFoot}{|v_k \protect\rangle}

\newcommand\trick[1]{}

\renewcommand{\[}{\begin{equation}}
\renewcommand{\]}{\end{equation}}

\begin{document}

\title{Analytic expressions for the steady-state current with finite extended reservoirs}
\author{Michael Zwolak}
\email{mpz@nist.gov}
\affiliation{Biophysical and Biomedical Measurement Group, Microsystems and Nanotechnology Division, Physical Measurement Laboratory, National Institute of Standards and Technology, Gaithersburg, MD 20899, USA}

\begin{abstract}
Open system simulations of quantum transport provide a platform for the study of true steady states, Floquet states, and the role of temperature, time-dynamics, and fluctuations, among other physical processes. They are rapidly gaining traction, especially techniques that revolve around ``extended reservoirs''---a collection of a finite number of degrees of freedom with relaxation that maintain a bias or temperature gradient---and have appeared under various guises (e.g., the extended or mesoscopic reservoir, auxiliary master equation, and driven Liouville--von Neumann approaches). Yet, there are still a number of open questions regarding the behavior and convergence of these techniques. Here, we derive general analytical solutions, and associated asymptotic analyses, for the steady-state current driven by finite reservoirs with proportional coupling to the system/junction. In doing so, we present a simplified and unified derivation of the non-interacting and many-body steady-state currents through arbitrary junctions, including outside of proportional coupling. We conjecture that the analytic solution for proportional coupling is the most general of its form for isomodal relaxation (i.e., relaxing proportional coupling will remove the ability to find compact, general analytical expressions for finite reservoirs). These results should be of broad utility in diagnosing the behavior and implementation of extended reservoir and related approaches, including the convergence to the Landauer limit (for non-interacting systems) and the Meir-Wingreen formula (for many-body systems). 
\end{abstract}

\maketitle

\section{Introduction}
\label{sec:Intro}

Nanoscale and molecular electronics encompasses a broad range of fundamental studies~\cite{zhao_cross-plane_2020,el_abbassi_robust_2019,yuan_transition_2018,zhou_photoconductance_2018} and applications, such as sensing in aqueous solution~\cite{zwolak_electronic_2005, zwolak_colloquium_2008, tsutsui_identifying_2010, huang_identifying_2010, chang_electronic_2010,zwolak_dna_2012, ohshiro_single-molecule_2012, li_detection_2018}. Sensing in inhomgeneous, dynamical environments, in particular, requires both statistical averaging~\cite{lagerqvist_fast_2006,lagerqvist_comment_2007,lagerqvist_influence_2007,gruss_graphene_2018,ochoa_generalized_2019,ochoa_optimal_2020} and the incorporation of dephasing~\cite{krems_effect_2009} and other quantum effects due to rapid atomic fluctuations---and potentially quite large energetic changes, e.g., due to ionic motion. The incorporation of fluctuations and quantum effects into simulation is challenging as it often requires the inclusion of additional degrees of freedom, such as explicit electronic reservoirs or phonon/vibrational bath modes. 

Along these lines, many transport processes can be treated within the so-called ``microcanonical approach,'' where a closed, finite system (including ``reservoirs'') are directly simulated~\cite{ventra_transport_2004, bushong_approach_2005, sai_microscopic_2007, chien_bosonic_2012, chien_interaction-induced_2013, chien_landauer_2014, gruss_energy-resolved_2018,zwolak_communication_2018}. There have been various implementations of finite, closed reservoirs within Matrix Product States (MPS)~\cite{cazalilla_time-dependent_2002,zwolak_mixed-state_2004,bohr_dmrg_2005,gobert_real-time_2005,bohr_strong_2007,al-hassanieh_adaptive_2006,schneider_conductance_2006,schmitteckert_signal_2006,dias_da_silva_transport_2008,heidrich-meisner_real-time_2009,branschadel_conductance_2010,chien_interaction-induced_2013,gruss_energy-resolved_2018}. These are computationally limited due to the rapid generation of entanglement during particle flow and scattering, but they can incorporate correlated impurity centers. Recent many-body implementations include a combined numerical-renormalization group and MPS approach~\cite{schwarz_nonequilibrium_2018}, as well as a technique that transforms the canonical basis---a basis that respects the scattering nature of transport and ``breaks'' the entanglement barrier---and exponentially improves simulation efficiency~\cite{rams_breaking_2020}.

Steady-states can rapidly form---with a rise time inversely proportional to the electronic bandwidth in certain cases~\cite{zwolak_communication_2018}. The microcanonical approach can thus still be quite powerful. However, statistical averaging over noise and fluctuations is computationally demanding since closed systems inevitably have recurrences and thus many separate simulations are required. Scanning over a parameter, such as coupling or temperature (e.g., included within a ``microcanonical'' picture via purification), can suffer from the same issue, requiring many independent simulations of a real-time evolution into a quasi-steady state. Quantum noise, dephasing, and periodic driving (Floquet states) present further issues, both demanding very large reservoirs or baths to act as effective sinks~\footnote{Whether found via closed or open systems, the range of periodic driving for a Floquet state will be limited, but in different ways.}. Opening up the reservoirs, via relaxation, can extend the scope of numerical simulations of nanoscale devices and sensors, especially the treatment of electronics in noisy environments or in the presence of quantum processes. Ultimately, the efficiency for any given task has to be directly assessed and compared between methods. For instance, within tensor networks, stochastic wave function techniques~\cite{transchel_monte_2014} or averaging over pure states~\cite{berta_thermal_2018} may be more efficient at solving the open system equations or handling finite temperature. More broadly, different computational machinery and approaches introduce a range of considerations that factor into the computational cost and accuracy, which will {\em not} be our objective here to assess. Rather, we will examine theoretical aspects of a class of open system simulations.

There have been a number of similar proposals for open-system simulations with explicit degrees of freedom to represent the reservoirs, such as ``extended reservoirs''~\cite{gruss_landauers_2016,elenewski_communication_2017,gruss_communication_2017}, the auxiliary quantum master equation approach (AMEA) which permits non-isomodal relaxation in the extended reservoirs~\cite{arrigoni_nonequilibrium_2013, dorda_auxiliary_2014, dorda_auxiliary_2015, dorda_optimized_2017, chen_auxiliary_2019,chen_markovian_2019}, and the driven Liouville--von Neumann  (DLvN) that treats non-interacting systems~\cite{zelovich_state_2014, zelovich_moleculelead_2015, zelovich_driven_2016, hod_driven_2016, zelovich_parameter-free_2017, chiang_quantum_2020}, as well as other approaches~\cite{sanchez_molecular_2006,subotnik_nonequilibrium_2009,dzhioev_super-fermion_2011} (some of which do not obey proper quantum evolution, see the overview at Ref.~\onlinecite{elenewski_communication_2017}). This particular class of open-system approaches to transport is in addition to a variety of other open-system techniques for transport and dynamics~\cite{kohn_quantum_1957,frensley_simulation_1985,frensley_boundary_1990,mizuta_transient_1991,fischetti_theory_1998,fischetti_master-equation_1999,knezevic_decoherence_2008,schaller_systematic_2009,zedler_weak-coupling_2009,breuer_stochastic_2009,jones_theoretical_2011,strumpfer_open_2012,knezevic_time-dependent_2013,rosati_microscopic_2014,schaller_relaxation_2014,yan_theory_2014,rosati_derivation_2014,rosati_microscopic_2015,purkayastha_out--equilibrium_2016,fischetti_overview_2016,rosati_lindblad_2017,purkayastha_quantum_2017,hartmann_exact_2017,schinabeck_hierarchical_2020,tanimura_numerically_2020}, which employ different strategies and, in some cases, apply to particular parameter ranges (e.g., weak coupling).

Explicit or extended reservoir approaches (ERAs) are all part of a general idea: A finite reservoir can effectively be transformed into a continuum by broadening the modes via relaxation~\cite{imamoglu_stochastic_1994,garraway_decay_1997,garraway_nonperturbative_1997,zwolak_dynamics_2008}, with recent rigorous results for bosonic systems as thermal quantum environments~\cite{tamascelli_nonperturbative_2018,tamascelli_efficient_2019,mascherpa_optimized_2020}. A related open-system approach was developed for classical thermal transport through DNA~\cite{velizhanin_driving_2011, chien_tunable_2013} and topological systems~\cite{chien_thermal_2017, chien_topological_2018}. The theoretical treatment and behavior of this approach follows Kramers turnover for friction-controlled rates~\cite{velizhanin_crossover_2015}, but applied at the simulation level rather than for condensed-phase chemical reaction rates~\cite{kramers_brownian_1940,hanggi_reaction-rate_1990}. The quantum case is analogous and also follows a Kramers turnover~\cite{gruss_landauers_2016,elenewski_communication_2017,gruss_communication_2017}, with only the intermediate relaxation regime giving currents commensurate with physical expectations (i.e., the relaxation-free Landauer result for non-interacting systems and the Meir-Wingreen formula for interacting systems).

Very recently, these open system ideas have been translated to novel tensor networks approaches. In Ref.~\onlinecite{wojtowicz_open-system_2020}, an open-system, time-dependent variational principle approach was taken to solve for the dynamics and steady states within the ``mixed basis'' (the basis that breaks the entanglement barrier in closed system transport by properly accounting for the scattering nature of current-carrying particles~\cite{rams_breaking_2020}). A similar approach was developed in Ref.~\onlinecite{brenes_tensor-network_2020}, which uses just the energy basis and the time-evolving block decimation algorithm with swaps, and applies the technique to quantum thermal machines. In Ref.~\onlinecite{fugger_nonequilibrium_2020}, building on earlier work~\cite{dorda_auxiliary_2015}, the authors implement the AMEA approach within MPS with swaps of sites and a particular organization of fermionic modes, and apply the technique to quantum transport in the Anderson model with a power-law spectral function. Similarly, Ref.~\onlinecite{lotem_renormalized_2020} employs numerical-renormalization group like ideas (e.g., logarithmic discretization outside the bias window) and a reorganization of modes within an MPS approach, see also Ref.~\onlinecite{schwarz_lindblad-driven_2016} for related work applied to non-interacting systems.

ERAs are thus already mature enough to be applied to quantum transport. However, it is clear that there are many issues that remain, some of which may hinder the application of these approaches or inadvertently lead to spurious results. Here, we will further develop the analytic and mathematical basis for using extended reservoirs for transport. Specifically, we will derive analytical expressions, within the  ``proportional coupling'' scenario, for transport through non-interacting and many-body quantum impurities for arbitrary relaxation strength. We conjecture that these are the most general analytic results of this form when relaxation is isomodal. We also examine the asymptotic limits. All of these expressions should be helpful in assessing and validating numerical implementations. Furthermore, we give a unified and simplified derivation of the non-interacting and many-body solutions to the extended reservoir quantum master equation. This includes an alternative approach of treating the Markovian relaxation directly within the Keldysh formalism (where previously Green's functions were computed directly from the Markovian equation of motion). We will discuss how these approaches with Markovian relaxation limit---in both non-interacting and many-body cases---to the relaxation-free Landauer and Meir-Wingreen formulas, respectively (note that non-Markovian relaxation already obeys exactly a Landauer or Meir-Wingreen formula at finite relaxation).

The outline of this article is as follows: In Sec.~\ref{sec:Summary}, we provide a brief summary of the main results, specifically the expressions for the steady-state current for many-body and non-interacting systems, the analytical expressions for proportional coupling, and the asymptotic expressions in the same scenario. We will also note which equations appear twice in this paper so the correspondence is clear. In Sec.~\ref{sec:Background}, we provide the connection between the Lindblad master equation with relaxation for many-body systems and the driven Liouville-von Neumann equation (DLvN) for non-interacting systems. In Sec.~\ref{sec:Current}, we provide a unified, general solution for Markovian and non-Markovian relaxation, and many-body and non-interacting systems, for the steady-state current. In Sec.~\ref{sec:PC}, we examine the solution in proportional coupling and derive a fully analytic result for finite reservoirs. In Sec.~\ref{sec:Asymp}, we present the asymptotic analyses. These results and calculations give a comprehensive and, we hope, accessible treatment of the steady-state current  within extended reservoir approaches to quantum transport.

\section{Summary}
\label{sec:Summary}

We first summarize the main results. Some of the equations thus appear twice in this manuscript, for which we give the correspondence here. Details of all quantities are in subsequent sections. 

We examine transport driven through a junction/impurity/system $\qs$ by a chemical potential or temperature drop between the left ($\ql$) and right ($\qr$) reservoirs, both of which are assumed to be non-interacting with a quadratic coupling to the system. Following the ERA solution developed in Ref.~\onlinecite{gruss_landauers_2016} (including averaging the current from the left and right reservoirs to create a symmetrized version 
of the current), the steady-state current, $I$, can always be written as 
\begin{align}
I = & \frac{\im e}{2} \int\frac{d\omega}{2\pi}\, \tr \left[ \left\{ \bGa^\ql-\bGa^\qr \right\} \bG^< \right. \notag \\
     & \left. + \left\{ \tilde{\bGa}^\ql-\tilde{\bGa}^\qr \right\} \left\{ \bG^{r} - \bG^{a} \right\} \right] . \label{eq:Curr_Sum}
\end{align}
This equation has a similar structure to the well-known Meir-Wingreen formula~\cite{meir_landauer_1992,jauho_time-dependent_1994}, being in terms of $\qs$'s Green's functions ($\bG^<, \bG^{r(a)}$) and reservoir spectral densities ($\bGa^{\ql(\qr)},\tilde{\bGa}^{\ql(\qr)}$). Crucially, however, it is distinct: $\tilde{\bGa}^{\ql(\qr)}$ have occupation factors that cannot be disentangled from the actual spectral density when Markovian relaxation is present, which has important consequences discussed throughout this work. We emphasize this equation is valid for both non-interacting and many-body impurities, as well as Markovian or non-Markovian relaxation, and whether there is proportional coupling or not. Equation~\eqref{eq:Curr_Sum} here is the same as Eq.~\eqref{eq:Curr} in Sec.~\ref{sec:Current}. The quantities appearing in Eq.~\eqref{eq:Curr_Sum} are the electron charge $e$, the weighted and unweighted spectral densities ($\bGa^\alpha$ and $\tilde{\bGa}^\alpha$, respectively) for reservoir $\alpha=\ql,\qr$, the system's full lesser Green's function $\bG^<$, and the system's full advanced and retarded Green's functions ($\bG^{r}$ and $\bG^{a}$, respectively). These are all $N_\qs \times N_\qs$ matrices representing correlations between the $N_\qs$ system modes. 

We briefly note here that the many-body current, Eq.~\eqref{eq:Curr_Sum} limits to the normal Meir-Wingreen formula as the reservoir size goes to infinity and then the relaxation strength goes to zero. That is, whether many-body or non-interacting, the exact result for the current in the standard---relaxation-free---Meir-Wingreen setup will result. This also entails that the normal Landauer formula will result in this limit for non-interacting electrons~\cite{gruss_landauers_2016}. This will be discussed in Sec.~\ref{sec:Lim}, see also Ref.~\cite{zwolak_comment_2020}.

For non-interacting systems, Eq.~\eqref{eq:Curr_Sum} becomes
\[ \label{eq:nonintCurr_Sum}
I=e \int\frac{d\omega}{2\pi} \tr \left[ \tilde{\bGa}^\ql \bG^a \bGa^\qr \bG^r - \bGa^\ql \bG^r \tilde{\bGa}^\qr \bG^a \right] .
\]
The Markovian nature of this equation is reflected in the appearance of two similar terms rather than a single term multiplied by a difference in Fermi-Dirac distributions. Eq.~\eqref{eq:nonintCurr_Sum} is the same as Eq.~\eqref{eq:nonintCurr} in Sec.~\ref{sec:nonintlimit}.

For ``proportional coupling,'' the two reservoirs have identical mode distributions and couplings (up to a proportionality factor) to the system modes (i.e., $\bGa^\qr = \lambda \bGa^\ql$). Since the reservoirs are finite, this entails that the reservoir modes have the same set of energies and relaxation strengths. Under this condition, one obtains
\[
I =   \frac{\im e \lambda}{1+\lambda} \int\frac{d\omega}{2\pi}\, \tr \left[ \Delta \tilde{\bGa} \left\{ \bG^{r} - \bG^{a} \right\} \right] . \label{eq:CurrPC_Sum}
\]
Equation~\eqref{eq:CurrPC_Sum} here is the same as Eq.~\eqref{eq:CurrPC} in Sec.~\ref{sec:PC}. The difference in the weighted spectral density is 
\[
\tilde{\bGa}^\ql - \tilde{\bGa}^\qr = \im \sum_{k\in \ql} \left(\tilde{f}_k^\ql - \tilde{f}_k^\qr \right) \left[g_{k}^{r}(\omega) - g_{k}^{a}(\omega) \right] \ok ,
\]
where the $\tilde{f}_k^{\ql(\qr)}$ are the Fermi-Dirac occupations evaluated at frequency $\omega_k$ ($\omega$) for Markovian (non-Markovian) relaxation and bias $\mu_{\ql(\qr)}$, $\inner{i}{v_k}=v_{ik}$ is the coupling between the mode $i\in\qs$ and the mode $k\in\ql$ (if $k\in\qr$, $v_{ik} \to \sqrt{\lambda} v_{ik}$), and the $g_{k}^{r(a)}=1/(\omega-\omega_k \pm \im \gamma_k/2)$ are the retarded (advanced) Green's functions for $k\in\ql$ (or $\qr$) with relaxation $\gamma_k>0$ but without contact to the system. The sum is over only the left reservoir since we take proportional coupling and the modes and relaxation are the same in the right reservoir. 

While we might not be able to solve for $\bG^{r(a)}$ due to many-body interactions or involved self-energies, we can still do the integration for Markovian relaxation, yielding 
\[ \label{eq:GenCurr_PC_Sum}
I = - \frac{2 e \lambda}{1+\lambda} \sum_{k\in \ql} \left(\tilde{f}_k^\ql - \tilde{f}_k^\qr \right) \bk\Im {\bG^r (\omega_k + \im \gamma_k/2 )} \kk ,
\]
which is analytic ($\bG^r$ is analytic in the upper half plane where it is evaluated here). Thus, the important assumption is to consider proportional coupling with Markovian relaxation. Non-Markovian relaxation, for instance, has the integrand dependent on the Fermi-Dirac distribution, and thus is not readily integrated. Equation~\eqref{eq:GenCurr_PC_Sum} here is the same as Eq.~\eqref{eq:GenCurr_PC} in Sec.~\ref{sec:PC}. 

We conjecture that Eq.~\eqref{eq:GenCurr_PC_Sum} is the most general analytic result of this form for two reservoir transport with isomodal relaxation---i.e., that relaxing the assumption of proportional coupling will, at best, result in only specific cases of finite reservoirs to be analytically solvable---since this derivation makes clear what is required: In order to perform the $\omega$ integration, we need that the integrand has functions that are only analytic in the upper or lower half plane. If this is not satisfied, for instance, due to the appearance of $\bG^< $, see Eq.~\eqref{eq:Curr_Sum}, the integral then relies on specific knowledge about the form of the system's Green's functions (opposed to just regions of analyticity). We will be surprised---but delighted---if this conjecture is not true. 

We can also analyze the asymptotic forms for Markovian relaxation, starting either from Eq.~\eqref{eq:CurrPC_Sum} or Eq.~\eqref{eq:GenCurr_PC_Sum}. For weak Markovian relaxation, we find 
\[ \label{eq:Curr_WG_PC_Sum}
I \approx \frac{2e \lambda}{(1+\lambda)^2} \sum_{k \in \ql} \gamma_k (\tilde{f}_k^\ql-\tilde{f}_k^\qr) .
\]
This equation is only derived for non-interacting systems, but allows for inhomogeneous $\gamma_k$ (where the inhomogeneity is across $k$, but not between $\ql$ and $\qr$). For strong relaxation, the current is 
\[ \label{eq:Curr_SG_PC_Sum}
I \approx \frac{4e \lambda}{1+\lambda} \sum_{k \in \ql} \frac{\ik}{\gamma_k} (\tilde{f}_k^\ql-\tilde{f}_k^\qr)  .
\]
This expression holds for both non-interacting and many-body systems and homogeneous or inhomogeneous relaxation. Equations~\eqref{eq:Curr_WG_PC_Sum} and~\eqref{eq:Curr_SG_PC_Sum} are the same as Eq.~\eqref{eq:Curr_WG_PC} and Eq.~\eqref{eq:Curr_SG_PC}, respectively, in Sec.~\ref{sec:Asymp}.

The set of these equations, Eqs.~\eqref{eq:GenCurr_PC_Sum},~\eqref{eq:Curr_WG_PC_Sum}, and~\eqref{eq:Curr_SG_PC_Sum}, give compact analytic expressions to assess the behavior, convergence, performance, and numerical implementation of the extended and auxiliary reservoir approaches for many-body or non-interacting systems, as well as the DLvN for non-interacting systems. 

\section{Background}
\label{sec:Background}

We first  define the equations we solve, since there are alternative forms in the literature. We start with a Lindblad master equation for the full (potentially many-body) density matrix, 
\begin{align} \label{eq:fullMaster}
\dot{\rho} = - \frac{\im}{\hbar} [H, \rho]
    &+ \sum_{k\in\ql\qr} \gamma_{k+} \left( \cid{k} \rho \ci{k}
        - \frac{1}{2} \left \{ \ci{k} \cid{k}, \rho\right \}\right) \notag \\
    &+ \sum_{k\in\ql\qr} \gamma_{k-} \left( \ci{k} \rho \cid{k}
        - \frac{1}{2} \left \{ \cid{k} \ci{k}, \rho \right \} \right) .
\end{align}
The first term in the master equation gives the Hamiltonian evolution of $\rho$ under $H$. The next two terms, which both contain the anticommutator $\left\{ A, B \right\}$, give the particle injection (depletion) into the reservoir states $k$ at a rate $\gamma_{k+}$ ($\gamma_{k-}$). The total Hamiltonian is 
\[ \label{eq:TotalHam}
H = \sum_{k\in \ql\qr} \hbar\omega_k \cid{k}\ci{k}
    + \sum_{k\in \ql\qr} \sum_{i \in \qs} \hbar\left(v_{ki}\cid{k}\ci{i}
                                                  + v_{ik}\cid{i}\ci{k} \right)
    + H_{\qs} ,
\]
where $\omega_k$ is the frequency of reservoir mode $k \in \ql\qr$, $v_{ik}=v_{ki}^\star$ is the coupling of that mode to the system $\qs$ at site $i$, and $\hbar$ is the reduced Planck's constant. The Hamiltonian $H_\qs$ is for $\qs$, which can have many-body interactions that the DLvN does not, and can not, include (beyond mean field). The operators $\cid{m}$ ($\ci{m}$) are the creation (annihilation) operators for the state $m \in \ql\qs\qr$. The index $m$ carries a numerical index, as well as all labels (electronic state, spin, reservoir or system mode).

Equation~\eqref{eq:TotalHam}, when the continuum limit is taken, is the standard starting point for transport through impurities, where non-interacting left ($\ql$) and right ($\qr$) reservoirs drive a current through a system $\qs$ via an applied potential or temperature drop. The reservoirs are connected to the system only via quadratic (hopping) terms and are not connected to each other.

As usual, one wants the relaxation terms to give an equilibrium state of the reservoirs in the absence of $\qs$. This requires that $\gamma_{k+} \equiv \gamma_k f_\alpha (\omega_k)$ and $\gamma_{k-} \equiv \gamma_k [1 - f_\alpha (\omega_k)]$, where $f_\alpha (\omega_k)$ is the Fermi-Dirac distribution in the $\alpha \in \{\ql,\qr\}$ reservoir at different chemical potentials or temperatures. One important point is that this equilibrium is actually a pseudo-equilibrium for the reservoirs that does not properly incorporate the broadening of the reservoirs' states due to the relaxation, which can result in, e.g., zero bias currents. This clearly unphysical situation has been investigated in Refs.~\onlinecite{gruss_landauers_2016,elenewski_communication_2017}. When there is symmetry between the left and right reservoir modes, the zero-bias anomaly goes to zero~\cite{gruss_landauers_2016}. However, there is still a related anomaly due to smearing of full states above the Fermi level and vice versa for empty~\cite{wojtowicz_dual_nodate}. 

Throughout this work, when we refer to Markovian relaxation, we are referring to Eq.~\eqref{eq:fullMaster}. When we refer to non-Markovian relaxation, we are referring to a similar relaxation that gives identical retarded and advanced Green's functions, but a proper Fermi level. We do not write its equation of motion (as its name implies, it has memory and is not a time-local equation when the external environments are integrated out). 

The AMEA approach has the same equation as shown in Eq.~\eqref{eq:fullMaster}. However, it allows for transitions between $k$ and $k^\p$ in the relaxation. We will limit ourselves here to energetically local (isomodal) relaxation ($k=k^\p$)~\footnote{For higher dimensions, momentum can be the index.}, but many results still hold for this more general situation. Moreover, in Ref.~\onlinecite{elenewski_communication_2017}, it was shown that there is an exact correspondence between the DLvN equation and the Lindblad master equation when electrons are taken to be non-interacting. Equation~\eqref{eq:fullMaster}, however, is more general since $\rho$ is the full, many-body density matrix on the exponentially large Hilbert space of the system $\qs$ and the left (right) reservoirs $\ql$ ($\qr$), and thus this can include many-body interactions (for many-body implementations of Eq.~\eqref{eq:fullMaster}, see Refs.~\onlinecite{wojtowicz_open-system_2020,brenes_tensor-network_2020,fugger_nonequilibrium_2020,lotem_renormalized_2020}). We emphasize that the state, $\rho$, is not the single-particle density matrix appearing in the DLvN, for which we reserve the $N \times N$ single-particle correlation matrix $\qc$ (this is related to a single-particle density matrix by normalizing the trace) where $N$ is the total number of electronic levels in $\ql\qs\qr$ (treating spin, if present, as a separate level). 

To make the correspondence of Eq.~\eqref{eq:fullMaster} with the DLvN, one has to take all regions to be non-interacting rather than just the reservoirs (and the reservoir-system coupling). When $H_\qs$ is also a non-interacting Hamiltonian, it can be written as
\begin{equation} \label{eq:nonintHS}
H_\qs=\sum_{i,j \in \qs} \bar{H}_{ij} \cid{i} \ci{j},
\end{equation}
where $\bar{H}_{ij}$ is the single-particle Hamiltonian. Note that we restrict $i,j$ to the system sites. When summing over all sites (the system and reservoirs), this gives the global single-particle Hamiltonian $H=\sum_{m,n} \bar{H}_{mn} \cid{m} \ci{n}$ with $m,n \in \ql \qs \qr$, which we will also use with the same symbol $\bH$ since the system single-particle Hamiltonian is just a submatrix of this larger matrix. These matrices are operators, but on the single-particle space rather than the full Hilbert space. 

With this introduction, it is now a simple matter to connect Eq.~\eqref{eq:fullMaster} to the DLvN~\cite{elenewski_communication_2017}. Assuming all regions to be non-interacting, one writes the equation of motion for the single-particle correlation matrix 
\[
\qc_{nm} = \tr \, [ \cid{m} \ci{n} \rho]
\]
with $m,n \in \ql \qs \qr$, which gives
\begin{equation} \label{eq:C}
\dot{\qc} = -\im [\bH,\qc]/\hbar + R[\qc],
\end{equation}
where $R[\qc]$ is the relaxation. In block form, as the DLvN uses, this equation has terms
\begin{equation}
\qc = \left( \begin{array}{ccc}
    \qc_{\mathcal{L,L}} & \qc_{\mathcal{L,S}} & \qc_{\mathcal{L,R}} \\
    \qc_{\mathcal{S,L}} & \qc_{\mathcal{S,S}} & \qc_{\mathcal{S,R}} \\
    \qc_{\mathcal{R,L}} & \qc_{\mathcal{R,S}} & \qc_{\mathcal{R,R}}
\end{array} \right) 
\end{equation}
and
\begin{equation} \label{eq:R}
R [\qc] = -\gamma \left(
\begin{array}{ccc}
    (\qc_{\mathcal{L,L}} - \qc_{0}^{\ql}) & \frac{1}{2} \qc_{\mathcal{L,S}}
        & \qc_\mathcal{{L,R}} \\
    \frac{1}{2} \qc_{\mathcal{S,L}} & 0 & \frac{1}{2} \qc_{\mathcal{S,R}} \\
    \qc_{\mathcal{R,L}} & \frac{1}{2} \qc_{\mathcal{R,S}}
        & (\qc_{\mathcal{R,R}} - \qc^\qr_{0})
\end{array}
\right) .
\end{equation}
The $\qc_{\alpha , \alpha^\p}$ are for a subset of states, i.e., in the regions $\alpha,\alpha^\p \in \{\ql, \qs, \qr\}$, where we have also taken a uniform $\gamma$ as otherwise one would have to include an inhomogeneous $\gamma$ into the blocks. This is a trivial matter but would hinder the direct comparison with the normal DLvN. However, for completeness, when inhomogeneous relaxation is present, the elements are
\[
(R[\qc])_{mn} = \gamma_{m+} \delta_{mn} -
        \frac{\qc_{mn}}{2}\left(\gamma_{m+} + \gamma_{m-}
                                + \gamma_{n_+} +\gamma_{n-}\right) ,
\]
where $m,n \in \ql \qs \qr$ and $\gamma_{m \pm} = 0$ if $m \in \qs$. 
The ``relaxed'' distributions are $\qc^{\alpha}_0 = \text{diag}[f_\alpha (\omega_k)]$. The connection between the Lindblad master equation, Eq.~\eqref{eq:fullMaster}, and Eq.~\eqref{eq:C} also shows that the DLvN is always a proper quantum evolution---that is, correlation matrix evolution is from a completely-positive trace-preserving map on the full density matrix. This proof works for finite or infinite reservoirs, unlike the result in Ref.~\onlinecite{hod_driven_2016} which holds only in the limit of infinite reservoirs. Moreover, since the Lindblad operators are in second quantized form, the evolution will always respect Pauli exclusion (observed empirically in Ref.~\onlinecite{zelovich_state_2014}). Further details are contained in Ref.~\onlinecite{elenewski_communication_2017}.

This connection entails that any solution of Eq.~\eqref{eq:fullMaster} when electrons are non-interacting is also a solution of the DLvN, Eq.~\eqref{eq:C}. As well, the solution will also be to the AMEA with only isomodal relaxation. We will start by recalling the computation for the current from Ref.~\onlinecite{gruss_landauers_2016}. Our presentation will differ from Ref.~\onlinecite{gruss_landauers_2016} only in two significant respects: We will give a simultaneous, unified approach for both the non-Markovian and Markovian extended reservoirs, including for many-body and non-interacting impurities. We will also assume the wide-band limit from the beginning. 
Since external environments, whether Markovian or non-Markovian, relax the reservoir modes, a true steady state exists even for finite reservoirs.  

We will need the lesser Green's function
\[ \label{eq:LessG}
\bG^<_{nm} (t,t^\p) = \im \avg{c^\dg_m(t^\p) c_n (t)},
\]
where the time-dependence signifies Heisenberg picture operators on the global space ($\ql \qs \qr$ plus external, implicit environments). This Green's function is connected to the correlation matrix elements through 
\[ \label{eq:CorrEleRel1}
 \qc_{nm} = -\im \bG^<_{nm} (t,t) .
 \]
 We will also need the retarded and advanced Green's functions,
\[ \label{eq:RetG}
\bG^r_{nm} (t,t^\p) = -\im \Theta(t-t^\p) \avg{ \left\{ c^\dg_m(t^\p) , c_n (t) \right\} }
\]
and
\[ \label{eq:AdvG}
\bG^a_{nm} (t,t^\p) = \im \Theta(t^\p-t) \avg{ \left\{ c^\dg_m(t^\p) , c_n (t) \right\} } ,
\]
respectively. Again, the $\left\{ A, B \right\}$ gives the anticommutator. Equations~\eqref{eq:RetG} and~\eqref{eq:AdvG} are related by $\left[ \bG^a_{nm} (t,t^\p) \right]^\star =  \bG^r_{mn} (t^\p,t)$. Within a stationary state, these Green's functions depend only on the time difference and, for the retarded and advanced Green's functions, we have $\bG^r (\omega)=\left[ \bG^a (\omega) \right]^\dg$ for their Fourier transforms. 

\section{Current}
\label{sec:Current}

By taking the derivative of the total particle number in the left or right reservoir, the time-dependent current is given by the sum over system-reservoir correlations. For the current from the left into the system, this reads
\[
I_\ql = \im e \sum_{k\in\ql,j\in\qs} \left( v_{kj} \avg{c_k^\dg c_j} - v_{jk} \avg{c_j^\dg c_k} \right)
\]
evaluated at time $t$. The current is related to the lesser Green's functions through
\[
I(t) = e \sum_{k\in \ql} \sum_{j \in \qs}
          \left[v_{kj} \bG_{jk}^{<}(t,t) - v_{jk} \bG_{kj}^{<}(t,t) \right] .
\]
Within the steady state, we take the Fourier transform
\[ \label{eq:Jless}
I = e \sum_{k\in \ql} \sum_{j \in \qs} \int\frac{d\omega}{2\pi} \,
       \left[v_{kj} \bG_{jk}^{<}(\omega) - v_{jk} \bG_{kj}^{<}(\omega) \right] .
\]
Given the non-interacting nature of the reservoirs, the Dyson equations for these Green's functions are
\[ \label{eq:Dysonjk}
\bG_{jk}^{<}(\omega) = \sum_{i \in \qs}v_{ik}
      \left[\bG_{ji}^{r}(\omega) g_{k}^{<}(\omega)
            + \bG_{ji}^{<}(\omega) g_{k}^{a}(\omega) \right] 
\]
and
\[ \label{eq:Dysonkj}
\bG_{kj}^{<}(\omega) = \sum_{i\in \qs} v_{ki}
     \left[g_{k}^{r}(\omega) \bG_{ij}^{<}(\omega)
           + g_{k}^{<}(\omega) \bG_{ij}^{a}(\omega) \right] .
\]
These stationary state expressions are found by the equation-of-motion method, which we provide in Appendix~\ref{sec:EOM} for the readers' convenience. 

The current, Eq.~\eqref{eq:Jless}, is the inverse Fourier transform at equal time $t-t^\p=0$ and can then be rewritten as
\begin{align} 
I = e \sum_{k\in \ql}\sum_{i,j \in \qs}\int\frac{d\omega}{2\pi}\, & v_{jk} v_{ki}
    \left\{ g_{k}^{<}(\omega)
           \left[\bG_{ij}^{r}(\omega)
                 - \bG_{ij}^{a}(\omega)\right] \right. \notag \\
    &\left. -\left[g_{k}^{r}(\omega) - g_{k}^{a}(\omega) \right]
           \bG_{ij}^{<}(\omega) \right\} . \label{eq:CurrL}
\end{align}
This expression applies to both the Markovian and non-Markovian reservoirs. Already, one can see here the appearance of self-energies $\bS_{ji} = \sum_k v_{jk} v_{ki} g_k$, with $g_k = g_k^<,g_k^{r(a)}$. The lesser components will depend non-Markovian or Markovian nature of relaxation.

To simplify the expression further while capturing both the Markovian and non-Markovian extended reservoirs, we define a new function
\[
\tilde{f}_k=\begin{cases}
f_{\alpha} \left(\omega\right) & \textbf{non--Markovian}\\
f_{\alpha} \left(\omega_{k}\right) & \textbf{Markovian} 
\end{cases} ,
\]
with $f_{\alpha} ( \omega)$ the normal Fermi-Dirac function with bias $\mu_{\alpha}$ and $\alpha = \ql (\qr)$ is the reservoir component of index $k$. This sets the occupation either according to the physical, non-Markovian equilibrium or the unphysical, Markovian equilibrium. In both situations, the Fermi-Dirac distribution $f$ appears, but either with the frequency appearing in the Green's function, $\omega$, or with the reservoir's isolated mode frequency, $\omega_k$. As we discuss extensively elsewhere~\cite{gruss_landauers_2016,elenewski_communication_2017}, the former correctly occupies the broadened mode, whereas the latter occupies and then broadens. As will be clear below, for weak external relaxation, these are approximately equal to each other, but diverge for strong relaxation resulting in unphysical---sometimes quite large and potentially zero-bias---currents.

Consolidating the terms by using the Keldysh equation 
\[ \label{eq:Keldyshk}
g_{k}^{<}(\omega)=g_k^r(\omega) \Sigma_k^<(\omega)  g_k^a(\omega)  ,
\]
where, for this non-interacting, ``isolated'' mode, 
\[ \label{eq:NM_Ek}
\begin{gathered}
\mathbf{non–Markovian} \\
\Sigma_k^<=\im \gamma_k f_\alpha (\omega)
\end{gathered}
\]
for the wide-band, non-Markovian extended reservoirs~\cite{gruss_landauers_2016} (herein called just ``non-Markovian'') and 
\[ \label{eq:M_Ek}
\begin{gathered}
\mathbf{Markovian} \\
\Sigma_k^<=\im  \gamma_{k+} \cdot 1 + \im  \gamma_{k-} \cdot 0 = \im \gamma_k f_\alpha (\omega_k)
\end{gathered}
\]
for the Markovian extended reservoirs. Note that the non-Markovian version connects each reservoir mode to a single external environment with a well-defined chemical potential. Each of these external environments for $\ql$ have the same well-defined chemical potential $\mu_\ql$ and the same for $\qr$, $\mu_\qr$ (this follows the calculations in Ref.~\onlinecite{gruss_landauers_2016} and see this same reference for the expressions outside of the wide-band limit). It is non-Markovian due to a surface---the Fermi level---even though it has no other structure. The Fermi level is sufficient to introduce memory (for non-Markovian relaxation with additional structure in these environments, see the expressions in Ref.~\onlinecite{gruss_landauers_2016}). 

The Markovian case represents the reservoir mode in contact with separate full and empty wide-band reservoirs, but with different couplings, which results in a finite mode occupancy. This case has to be wide band, as otherwise it would not be Markovian. Note that, in Ref.~\onlinecite{gruss_landauers_2016}, we derived the Green's functions directly from the Lindblad equation of motion, Eq.~\eqref{eq:fullMaster}. Here, however, we simply take a Keldysh  approach and use the fact that the Markovian master equation, Eq.~\eqref{eq:fullMaster}, is exactly the equation of motion that one gets when each reservoir mode is connected to two reservoirs, one full and one empty giving the factors 1 (i.e., $f(\omega)=1$) and 0 (i.e., $f(\omega)=0$). The lack of a well-defined chemical potential (i.e., they are at $\pm \infty$) points to a serious issue that has repercussions in calculations that must be avoided by a careful examination of behavior versus parameters. 

Equation~\eqref{eq:Keldyshk} with Eq.~\eqref{eq:NM_Ek} or~\eqref{eq:M_Ek} yields the identity (for either the non-Markovian or Markovian reservoirs) 
\[ \label{eq:gl_identity}
g_{k}^{<}(\omega) = -\tilde{f}_k \left[g_{k}^{r}(\omega)-g_{k}^{a}(\omega)\right] .
\] 
Defining the weighted and unweighted spectral densities (i.e., weighted and unweighted versions of the normal $\Gamma^{\ql (\qr)} = \im ( \bS^r_{\ql (\qr)} - \bS^a_{\ql (\qr)}) = -2 \Im \bS^r_{\ql (\qr)}$) as 
\[  \label{eq:weightedsigma_gen}
\tilde{\bGa}^{\ql(\qr)}(\omega) = \im\sum_{k\in \ql(\qr)} \tilde{f}_k  \left[g_{k}^{r}(\omega) - g_{k}^{a}(\omega) \right] \ok
\]
and
\[  \label{eq:unweightedsigma_gen}
\bGa^{\ql(\qr)}(\omega) = \im\sum_{k\in \ql(\qr)}  \left[g_{k}^{r}(\omega) - g_{k}^{a}(\omega) \right] \ok ,
\]
respectively, the current, Eq.~\eqref{eq:CurrL}, becomes
\[ \label{eq:CurrLcom}
I = \im e \int\frac{d\omega}{2\pi}\, \tr \left[ \bGa^\ql \bG^< + \tilde{\bGa}^\ql \left\{ \bG^{r} - \bG^{a} \right\} \right]
\]
This is structurally similar but distinct from the Meir-Wingreen result (c.f., Eq.~(18) in Ref.~\cite{jauho_time-dependent_1994}). For Markovian relaxation, one cannot remove the occupation factor from $\tilde{\bGa}$. Thus, while analogous, it is different than their formula, providing the exact solution to transport through a many-body impurity according to Eq.~\eqref{eq:fullMaster}.

Moreover, we will apply it to finite reservoirs where each mode has a finite lifetime. In the wide-band limit (see Ref.~\onlinecite{gruss_landauers_2016} for the more general case), the Green's functions of the reservoir modes are the following: For both non-Markovian and Markovian extended reservoirs, the retarded (advanced) Green's functions are 
\[ \label{eq:singlegreen}
g_{k}^{r (a)}(\omega) = \frac{1}{\omega-\omega_{k} \pm \im\gamma_k/2} .
\]
The lesser Green's functions can be found by direct computation or using the Keldysh equation of the reservoir mode in contact with the external environment. We use the latter, $g^<_k = g_k^r \Sigma^<_k g_k^a = g_k^r \Sigma^<_k \left[g_k^r\right]^{*}$. This gives
\[ \label{eq:gl}
g_{k}^{<}(\omega) =\frac{\im\gamma_k \tilde{f}_k}{\left(\omega-\omega_{k}\right)^{2} + \gamma_k^{2}/4} .
\]
The issue of occupancy becomes clear with this equation: The non-Markovian equilibrium occupies every part of the single-mode density of states according to the Fermi-Dirac distribution, $\tilde{f}_k= f_\alpha(\omega)$. The Markovian equilibrium, though, occupies the whole spectrum according to its completely isolated frequency $\omega_k$, $\tilde{f}_k= f_\alpha(\omega_k)$, violating the fluctuation-dissipation theorem. As long as the relaxation strength, $\gamma_k$, is weak, there is little difference. This directly leads to the condition, $\gamma_k \ll k_B T/\hbar$, that the relaxation has to be much less than the thermal relaxation for the Markovian approximation to be valid~\cite{gruss_landauers_2016}. The condition, in some situations, is not so strict, as many (but not all) models will give the correct Landauer result (for non-interacting systems) even when this is not satisfied. This comes down to the fact that there is often a temperature scale in the problem, $T^\star$, below which the current becomes independent of temperature, see Sec.~\ref{sec:Lim}. When this is the case, a ``too large'' relaxation strength can still give the correct result. This is a heuristic---a rule-of-thumb for practical calculation---that should be employed with care.

We can give symmetric expressions by averaging the left and right currents,
\begin{align}
I = & \frac{\im e}{2} \int\frac{d\omega}{2\pi}\, \tr \left[ \left\{ \bGa^\ql-\bGa^\qr \right\} \bG^< \right. \notag \\
     & \left. + \left\{ \tilde{\bGa}^\ql-\tilde{\bGa}^\qr \right\} \left\{ \bG^{r} - \bG^{a} \right\} \right] . \label{eq:Curr}
\end{align}
This equation is valid for both non-interacting and many-body system Hamiltonians $H_\qs$, as well as for both non-Markovian and Markovian relaxation. Indeed, it is our starting point for all subsequent calculations and also appears as Eq.~\eqref{eq:Curr_Sum} in the Summary, Sec.~\ref{sec:Summary}.

Again, the difference with the standard approach of Meir-Wingreen~\cite{meir_landauer_1992,jauho_time-dependent_1994}---besides being for finite reservoirs---is the lack of ability to pull out the occupancies from the $\tilde{\bGa}$'s. For the non-Markovian reservoirs, the current vanishes when in equilibrium, $f_\ql (\omega)=f_\qr (\omega) \equiv f_{eq} (\omega)$ and $\bG^<=-f_{eq} (\bG^r-\bG^a)$. For the Markovian case, the current vanishes in equilibrium for symmetric reservoirs in a proportional coupling scenario. For non-proportional coupling, one can often have zero bias currents, pointing to one of the particular pathologies of the Markovian setup~\footnote{Non-proportional coupling can also lack the anomalous zero-bias current. The simplest case is a single-site impurity between single-mode reservoirs at $T=0$. When the $\ql$ and $\qr$ modes are each below the Fermi level, there is no current at zero bias (and indeed, there is zero current identically until the bias results in one mode above and another below the Fermi level. More pronounced is the case where the  $\ql$ and $\qr$ modes are at the same frequency but, to be non-proportional, have different relaxation strengths. This always has zero current. However, what this counter example shows is that the lack of a zero bias current is measure zero in parameter space. For unequal  $\ql$ and $\qr$ mode frequencies, any $T \neq 0$ has a zero bias current. For equal  $\ql$ and $\qr$ mode frequencies, the setup has already been sufficiently restricted and is of measure zero itself in parameter space. Proportional coupling, of course, is also measure zero in parameter space. We conjecture that the lack of zero-bias anomalous behavior only happens within scenarios of measure zero. Moreover, as we will show elsewhere, there is still anomalous behavior, but it happens in the presence of a bias~\cite{wojtowicz_dual_nodate}.}. These currents, though, are related to $\hbar \gamma_k/k_B T$ and vanish as this goes to zero. We caution that even in the proportional coupling scenario, the Markovian case still gives anomalous currents. We discuss these elsewhere~\cite{wojtowicz_dual_nodate} but they have the same mathematical origin: The Markovian equation occupies and then broadens, which results in unphysical occupied (and unoccupied) states at high and low energy. Essentially, one will have Lorentzian broadened ``Fermi level'' rather than a Fermi-Dirac exponential cutoff. 

From here, there are two special cases that are useful. One is to maintain generality and allow for many-body interactions in $\qs$, but assume proportional coupling in the reservoirs, and the other is to assume a globally non-interacting lattice (in the single particle sense, which is quite strict but widely applied for molecular and nanoscale electronic junctions, including in density functional theory approaches). We will first treat non-interacting systems and then outline the general (non-interacting and many-body) correspondence to Landauer and Meir-Wingreen formulas. In Sec.\ref{sec:PC}, we will examine proportional coupling.

\subsection{Non-interacting Systems}
\label{sec:nonintlimit}

For non-interacting electrons, one can also simplify the current in the standard way~\cite{meir_landauer_1992,jauho_time-dependent_1994,caroli_direct_1971}. Starting from Eq.~\eqref{eq:Curr}, one uses the Keldysh equation,
\[ \label{eq:nonintKeldysh}
\bG^< = \bG^r \bS^< \bG^a \to \im \bG^r (\tilde{\bGa}^\ql+\tilde{\bGa}^\qr) \bG^a
\]
and the identities 
\[ \label{eq:nonintId}
\bG^r-\bG^a = -\im \bG^r (\bGa^\ql + \bGa^\qr) \bG^a = -\im \bG^a (\bGa^\ql + \bGa^\qr) \bG^r ,
\]
which both employ non-interacting electrons (after the arrow for the first equation, as one can include the many-body lesser self-energy)~\cite{haug_quantum_2008}. Again, the Markovian extended reservoirs require that we retain the occupancy within the spectral density. Employing the rightmost expressions in Eqs.~\eqref{eq:nonintKeldysh} and~\eqref{eq:nonintId}, Eq.~\eqref{eq:Curr} becomes
\[ \label{eq:nonintCurr}
I=e \int\frac{d\omega}{2\pi} \tr \left[ \tilde{\bGa}^\ql \bG^a \bGa^\qr \bG^r - \bGa^\ql \bG^r \tilde{\bGa}^\qr \bG^a \right] .
\]
Equation~\eqref{eq:nonintCurr} is the exact solution to the Markovian equation of motion, Eq.~\eqref{eq:fullMaster}, when the impurity region is non-interacting and thus it is the exact solution to the driven Liouville--von Neumann approach, Eq.~\eqref{eq:C}, for arbitrary reservoirs and parameters.
When the relaxation is non-Markovian, the solution reduces to the more traditional expression
\[ \label{eq:nonintCurrStandard}
I=e \int\frac{d\omega}{2\pi} \left( f_\ql (\omega) - f_\qr (\omega) \right) \tr \left[ \bGa^\ql \bG^r \bGa^\qr \bG^a \right] ,
\]
since in this case there is a well-defined Fermi level~\footnote{Note that to get Eq.~\eqref{eq:nonintCurrStandard}, we also used that $\tr \left[ \tilde{\bGa}^\ql \bG^a \bGa^\qr \bG^r \right]$ is equal to $\tr \left[ \tilde{\bGa}^\ql \bG^r \bGa^\qr \bG^a \right]$. When relaxation is non-Markovian, this follows by taking $\bGa^\qr = \bGa^\qr + \bGa^\ql - \bGa^\ql$, applying Eq.~\eqref{eq:nonintId}, and canceling $\bGa^\ql$ terms by pulling out $f_\ql$ (which is allowed since this does not depend on $k$) and using the cyclic property of the trace. When relaxation is Markovian, the Fermi-Dirac distribution can not be pulled out of $\tilde{\bGa}^\ql$.}. Both of these expressions just require computing the retarded (advanced) Green's functions, $\bG^{r (a)} = 1/(\omega - \bHS - \bS^{r (a)})$ and self-energies $\bS^{r (a)}=\sum_{k\in\ql\qr} g^{r (a)}_k \ok$, and the weighted spectral density. We note here, of course, that in the non-interacting limit, one can compute transport properties in the normal way. The primary benefit of ERAs is that they provide a numerical framework for many-body systems, as well as time dynamics. Equation~\eqref{eq:nonintCurr}, though, can be used as a standard tool to assess behavior (such as the presence of anomalous currents~\cite{wojtowicz_dual_nodate}), evaluate performance (e.g., in using different reservoir discretizations~\cite{zwolak_finite_2008,elenewski_performance_nodate}), and validate numerical implementation of ERAs~\cite{wojtowicz_open-system_2020,wojtowicz_dual_nodate,elenewski_performance_nodate}.

\subsection{Landauer and Meir-Wingreen Correspondence}
\label{sec:Lim}

Whether one works with a many-body impurity, Eq.~\eqref{eq:Curr} generally or Eq.~\eqref{eq:CurrPC} for proportional coupling, or the non-interacting result, Eq.~\eqref{eq:nonintCurr} or Eq.~\eqref{eq:CurrPC} for proportional coupling, the Markovian expressions converge to non-Markovian expressions, as this has to do with the convergence of the weighted spectral density---alternatively, the lesser Green's function---of the finite reservoir.  This was examined in detail in Ref.~\onlinecite{gruss_landauers_2016}. There, it was shown that the one-norm difference of the Markovian and non-Markovian lesser Green's function is bounded by 
\[
\frac{\hbar \gamma_k}{4 k_B T} \ln \frac{k_B T}{\hbar \gamma_k} .
\]
That is, it is controlled by the ratio of the relaxation strength and the thermal energy. This also helps put bounds on the necessary reservoir size~\cite{gruss_landauers_2016,elenewski_communication_2017}. Clearly, as the temperature goes to zero, convergence requires a very small relaxation strength, as the derivative of the occupation at the Fermi level is diverging. However, while this can influence the accuracy of the calculation if there are sharp features at the Fermi level, e.g., due to interference effects~\cite{solomon_understanding_2008,hsu_single-molecule_2012,lambert_basic_2015,chiang_quantum_2020}, in many other cases one may not see its effect. For instance, as seen in Ref.~\onlinecite{wojtowicz_open-system_2020}, below a certain temperature, $T^\star$, which potentially could depend on the voltage for sharp features, the current is not changing~\footnote{This temperature is dependent on voltage and on features in the density of states. A smooth density of states (in both the reservoirs and the system) over the bias window increases $T^\star$, as can many-body relaxation.}. Thus, so long as $\gamma_k \ll  k_B T^\star/\hbar$, the current will still be accurate even though the lesser Green's function has not formally converged. As emphasized above, this has be applied with care.

When the weighted spectral density and lesser Green's functions have approximately converged to their non-Markovian counterparts, the current for Markovian relaxation approximately obeys a Landauer or Meir-Wingreen formula (for non-interacting and many-body interactions, respectively). It seems unlikely that an exact Landauer expression with relaxation will be obeyed unless the relaxation is both non-Markovian and given by a single-particle mechanism. 

The finite extended reservoir, whether Markovian or non-Markovian, will then converge to the relaxation-free infinite reservoir as $N \to \infty$ and then $\gamma_k \to 0$. As done in Ref.~\onlinecite{gruss_landauers_2016}, the limit to the Landauer formula can now be taken directly with Eq.~\eqref{eq:nonintCurr}. First one takes the infinite reservoir limit, then when $\gamma_k$ is sufficiently narrow in the sense already discussed, one can replace $\tilde{f}_k$ with $f(\omega)$ (for the Markovian case, since the non-Markovian already has this) in Eq.~\eqref{eq:weightedsigma_gen}.  This shows that the current in the steady-state of the Markovian master equation, Eq.~\eqref{eq:fullMaster}, limits to the Landauer result. It requires both that the reservoir goes to infinity (first) and then the relaxation strength to zero~\cite{gruss_landauers_2016,zwolak_comment_2020}.

For many-body systems, one has to work instead with Eq.~\eqref{eq:Curr}. The same replacement applies, noting that it occurs both in the weighted spectral density and within $\bG^<$. One can think of this process in an alternative manner: The lesser Green's function of the truly isolated reservoir is a delta function in frequency space and one is using instead the Lorentzian representation of the delta function (the retarded and advanced Green's functions are no different than normal, except one has the broadening tied to $\gamma_k/2$ rather than ``$\eta \to 0$'', with $\eta$ the control parameter appearing in the Green's functions whose sole purpose is to ensure causality---or, for what is important here, analyticity in the upper or lower half plane). In mathematical terms, it is expressing the following replacement:
\begin{align}
\lim_{\gamma_k \to 0} \im  \left[ g_{k}^{r}(\omega) - g_{k}^{a}(\omega) \right] \tilde{f}_\alpha & = 2 \pi \delta(\omega-\omega_k) \tilde{f}_\alpha \notag \\
& = 2 \pi \delta(\omega-\omega_k) f_\alpha(\omega) ,
\end{align}
which is exact as the relaxation strength goes to zero. As already noted, the reservoir size has to already be infinite before taking this limit~\footnote{If the reservoir size is finite, $\gamma_k \to 0$ will result in zero current. Nevertheless, it will still limit to a Landauer or Meir-Wingreen formula, as these formulas for non-Markovian relaxation are exact for all relaxation strengths and reservoir sizes. Hence, all that is required is that the Markovian relaxation expressions limit to the non-Markovian ones}.

\section{Proportional Coupling}
\label{sec:PC}

When the reservoirs are identical in density of states and distribution of couplings to system (allowing for the total coupling scale to be different), one obtains the proportional coupling case~\footnote{Note that we take $\bGa^\qr = \lambda \bGa^\ql$ rather than $\bGa^\ql = \lambda \bGa^\qr$, as is done in Refs.~\onlinecite{meir_landauer_1992,jauho_time-dependent_1994}, to maintain consistency with writing expressions using the left reservoir modes in Ref.~\onlinecite{gruss_landauers_2016}}, $\bGa^\qr = \lambda \bGa^\ql$, which allows one to eliminate the need to compute $\bG^<$. To exactly hold, e.g., in widely-employed tight-binding models, for instance, this would require all atoms bound to one of the electrodes to also be bound---up to the same proportionality constant---to the other electrode. This condition is quite limiting. It can be, however, approximately valid in some cases. There are, for example, situations where one or a few junction states dominate transport, such as a symmetrically bound HOMO (highest occupied molecular orbital) level. Projecting on the subspace of relevant modes would, in these particular cases, give identical (or proportional) coupling. We investigate proportional coupling here as a test case for numerical implementations, as well as to understand behavior and convergence to the relaxation-free, continuum (i.e., the exact) results.

Following Refs.~\onlinecite{meir_landauer_1992,jauho_time-dependent_1994}, we take the steady state current as a weighted average of left and right currents, $I=x I_\ql+ (1-x) I_\qr$ with $x\in \left[0,1\right]$. Employing $x=\lambda/(1+\lambda)$ eliminates $\bG^<$ from the expression for the current. However, unlike Refs.~\onlinecite{meir_landauer_1992,jauho_time-dependent_1994}, we have to separately address $\tilde{\bGa}$. Proportional coupling requires that the discrete set of modes in $\ql$ and $\qr$ are the same, as well as their relaxation (it can be inhomogeneous in $k$ but identical between modes $\omega_k$ in $\ql$ and $\qr$). The couplings to the system need to satisfy $v_{k^\p i}=\sqrt{\lambda} v_{ki} \delta_{kk^\p}$ with $k^\p \in \qr$, $k \in \ql$, and $\delta_{kk^\p}$ signifying that the numerical part of the index is identical. This entails
\begin{align}
x \tilde{\bGa}^{\ql} - (1-x) \tilde{\bGa}^{\qr} = \im \frac{\lambda}{1+\lambda} \sum_{k\in \ql} & \left(\tilde{f}_k^\ql - \tilde{f}_k^\qr \right) \ok  \notag \\
	& \times \left[g_{k}^{r}(\omega) - g_{k}^{a}(\omega) \right] , 
\end{align}
for which we define 
\begin{align}
x \tilde{\bGa}^{\ql} - (1-x) \tilde{\bGa}^{\qr} \equiv \frac{\lambda}{1+\lambda} \Delta \tilde{\bGa}
\end{align}
at $x=\lambda/(1+\lambda)$. The current can then be written as 
\[ \label{eq:CurrPC}
I = \im e  \frac{\lambda}{1+\lambda} \int\frac{d\omega}{2\pi}\, \tr \left[ \Delta \tilde{\bGa} \left\{ \bG^{r} - \bG^{a} \right\} \right] . 
\]
Equation~\eqref{eq:CurrPC} is the same as Eq.~\eqref{eq:CurrPC_Sum} and is the one employed to derive the general result for finite reservoirs for Markovian relaxation (see below). This equation applies to many-body and non-interacting systems with proportional coupling, as well as to non-Markovian and Markovian relaxation. Note that since $\Delta \tilde{\bGa}$ goes to zero as the bias goes to zero for proportionally coupled reservoirs with Markovian relaxation, the current is zero when no bias is present. For non-Markovian relaxation, the current is always zero when no bias is present, for both proportional and non-proportional coupling. Anomalies, however, can still be present for both cases, as we will discuss elsewhere~\cite{wojtowicz_dual_nodate}.

For non-Markovian relaxation, Eq.~\eqref{eq:CurrPC} is equivalent to the Meir-Wingreen expression (c.f., Eq.~(21) in Ref.~\onlinecite{jauho_time-dependent_1994}) as one can pull out the Fermi-Dirac distributions:
\[ \label{eq:Curre_NM_PC}
I = \im e  \frac{\lambda}{1+\lambda} \int\frac{d\omega}{2\pi}\, \left( f_\ql (\omega) - f_\qr (\omega) \right) \tr \left[ \bGa \left\{ \bG^{r} - \bG^{a} \right\} \right]
\]
In this non-Markovian case, this is a ``pseudo-Landauer'' form (note $\bGa=\bGa^\ql=\bGa^\qr/\lambda$) and it has a nice interpretation: It is the spectral overlap of the reservoir and  broadened system density of states within the bias window that determines the current. For Markovian extended reservoirs, though, one can not pull out the Fermi-Dirac distributions and this interpretation breaks down. There is still a notion of spectral overlap, but it is the spectral overlap weighted by the difference in the Fermi-Dirac distributions of the {\em isolated} reservoir states. This is not equivalent to the former and gives anomalous behavior that has to be accounted for in practical calculations~\cite{wojtowicz_dual_nodate}.  Unlike the Markovian case within Eq.~\eqref{eq:CurrPC}, Eq.~\eqref{eq:Curre_NM_PC} can not be analyzed directly since the Fermi-Dirac distributions depend on $\omega$.

When the relaxation is Markovian, one has a sum over $k\in\ql$ with each term having a product 
\[
\left[g_{k}^{r}(\omega) - g_{k}^{a}(\omega) \right]  \bk\left\{ \bG^r - \bG^a \right\} \kk ,
\]
where we used that the coupling matrix for each $k$ is an outer product in order to take the trace. This contains all the $\omega$-dependent factors. We can thus do integrations in either the upper or lower half-plane depending on the functions in the integrand. For instance, $g_{k}^{r} \bG^r$ can be integrated in the upper half-plane giving zero (the retarded functions are analytic there) and $g_{k}^{r} \bG^a$ can be integrated in the lower half-plane since we do not know necessarily where the poles of $\bG^a$ are (but they are all in the upper half-plane). The $\omega$ integral in Eq.~\eqref{eq:CurrPC} thus gives
\[ \label{eq:GenCurr_PC}
I = - \frac{2 e \lambda}{1+\lambda} \sum_{k\in \ql} \left(\tilde{f}_k^\ql - \tilde{f}_k^\qr \right) \bk\Im {\bG^r (\omega_k + \im \gamma_k/2 )} \kk .
\]
This derivation demonstrates that this is a very general equation. It assumes proportional coupling and Markovian relaxation but otherwise applies to many-body or non-interacting systems and homogeneous or inhomogeneous relaxation. For non-Markovian relaxation, one has to work with Eq.~\eqref{eq:Curre_NM_PC} instead since the Fermi-Dirac distribution depends on the integration variable. Equation~\eqref{eq:GenCurr_PC} is the same as Eq.~\eqref{eq:GenCurr_PC_Sum} in the summary, Sec.~\ref{sec:Summary}.

The derivation, though, also demonstrates why a successful analytic calculation results: To obtain the current, one need only to deal with various retarded and advanced Green's functions. These have analytic properties that, e.g., the lesser Green's function does not. Hence, this is why we conjecture that Eq.~\eqref{eq:GenCurr_PC} is the most general result of its form for isomodal relaxation.

\section{Asymptotic Analyses}
\label{sec:Asymp}

Taking Eq.~\eqref{eq:GenCurr_PC} for proportional coupling and further assuming that the system is non-interacting, one can directly take the small and large relaxation limits. When the system is non-interacting, the retarded Green's function is 
\[ \label{eq:GrNonint}
\bG^r = \frac{1}{\omega - \bHS - \bS^r}
\]
with
\begin{align}
\bS^r & =  \sum_{k \in \ql\qr} \frac{\ok}{\omega-\omega_k+\im \gamma_k/2} \label{eq:selfenergy_NI} \\
         & = (1+\lambda) \sum_{k \in \ql} \frac{\ok}{\omega-\omega_k+\im \gamma_k/2} 
         \label{eq:selfenergy_NI_PC}
\end{align}
where the second line is for proportional coupling and the factor of $(1+\lambda)$ is to account for both reservoirs when the sum goes over only $\ql$. 

\subsection{Weak Relaxation}
\label{sec:WR}

When the $\gamma_k$ are all much smaller than the level spacing, $\gamma_k \ll \omega_k - \omega_{k^\p} \, \forall \, k,k^\p$, the self-energy evaluated at $\omega_k + \im \gamma_k/2$ becomes
\[ \label{eq:SESG}
\bS^r (\omega_k + \im \gamma_k/2) \approx -\im (1+\lambda) \frac{\ok}{\gamma_k}.
\]
That is, for very small $\gamma_k$, the $k^\mathrm{th}$ term is picked out. Moreover, Eq.~\eqref{eq:SESG} diverges and we will have that 
\[
\left\| \bS^r (\omega_k + \im \gamma_k/2) \right\| \gg \left\| \omega_k + \im \gamma_k/2 - \bHS \right\|
\]
assuming all elements of $\bHS$ are bounded and we have taken the operator norm. The self-energy from mode $k$ is thus dominant in the Green's function. We can then either retain all elements of the Green's function and diagonalize it perturbatively or we can note that due to the $\bk\circ \kk$ in the numerator of Eq.~\eqref{eq:GenCurr_PC}, we only need to invert the Green's function on a subspace of rank one. That is, the coupling matrix for one mode $k$, $\ok$, is of rank 1 (the operator acts on the whole single-particle space of the system, and thus is an $N_\qs \times N_\qs$ matrix with $N_\qs$ levels in the system) and can be inverted on that subspace. When put into Eq.~\eqref{eq:GenCurr_PC}, this gives $\bk(\ok)^{-1} \kk = 1$ (where 1 is the identity matrix). 

For small $\gamma_k$, Eq.~\eqref{eq:GenCurr_PC} thus yields
\[ \label{eq:Curr_WG_PC}
I \approx \frac{2e \lambda}{(1+\lambda)^2} \sum_{k \in \ql} \gamma_k (\tilde{f}_k^\ql-\tilde{f}_k^\qr) ,
\]
While we provided an identical analytic expression in Ref.~\onlinecite{gruss_landauers_2016} for $\lambda=1$, this now gives an alternative demonstration that it applies to all non-interacting junctions in the proportional coupling scenario. It further generalizes the result to inhomogeneous $\gamma$. Many-body systems, though, do not seem to be easily treated with this approach for small $\gamma$. As the number of reservoir modes increases towards the continuum limit, we note that the region of validity of Eq.~\eqref{eq:Curr_WG_PC} is pushed towards smaller values of $\gamma_k$ (e.g., generically as the inverse of the number of modes). Equation~\eqref{eq:Curr_WG_PC} is Equation~\eqref{eq:Curr_WG_PC_Sum} in the Summary, Sec.~\ref{sec:Summary}.

\subsection{Strong Relaxation}
\label{sec:SR}

When the $\gamma_k$ are all much larger than any frequency scale in the problem, $\gamma_k \gg \left\| \bH \right\| \forall \, k \in \ql$ (or $k \in \qr$), the self-energy is small (it is bounded by a function of $1/\gamma_k$). In words, the system itself---just the junction---is effectively almost closed, as excitations take a long time to decay into the extended reservoirs (and out to the external environments). The denominator of the Green's function is dominated by the extra factor of $\gamma_k$ passed to it through its argument. Thus, 
\[
\bG^r (\omega_k + \im \gamma_k/2 ) \approx -\frac{2 \im}{\gamma_k} .
\]
Putting this into Eq.~\eqref{eq:GenCurr_PC}, gives 
\[ \label{eq:Curr_SG_PC}
I \approx \frac{4e \lambda}{1+\lambda} \sum_{k \in \ql} \frac{\ik}{\gamma_k} (\tilde{f}_k^\ql-\tilde{f}_k^\qr)  .
\]
This is exactly the expression already derived in Ref.~\onlinecite{gruss_landauers_2016}, albeit for homogeneous $\gamma_k$, which, however, was not essential in the derivation, and $\lambda=1$. Equation~\eqref{eq:Curr_SG_PC} is Equation~\eqref{eq:Curr_SG_PC_Sum} in the Summary, Sec.~\ref{sec:Summary}.

Equation~\eqref{eq:GenCurr_PC}, though, holds for many-body systems, as well. We can thus go further. The key expression~\footnote{This is Eq.~(B17) in Ref.~\onlinecite{gruss_landauers_2016} with the minor typo (a missing $t$ in the exponent) corrected.} transforms back to real-time, 
\begin{align} \label{eq:essGr}
\int_{-\infty}^{\infty} &\frac{ d\omega \; \gamma_k}
                      {(\omega - \omega_k)^2 + \gamma_k^2/4} \bG^r_{ij}(\omega) \\
    &= 2\pi\int_{-\infty}^{\infty} dt \, e^{-\im\omega_{k}t-\gamma_k|t|/2} \;
            \bG_{ij}^{r}(-t)
    \approx - \im \frac{4\pi}{\gamma_k} \delta_{ij} , \notag
\end{align}
where we can take the imaginary part after integration since the Lorentzian is real. The exponential cutoff due to the large $\gamma_k$ limits the contribution to the integral to only a short time, $1/\gamma_k$. Using Eq.~\eqref{eq:RetG}, the short time approximation is simply the identity times $-\im$, but only for positive arguments. Also note that the Green's function is for the system, so that the self-energy from the reservoirs is small and smeared at large $\gamma$ (that is, large $\gamma$ does not introduce a fast process into $\bG^r$ even though Eq.~\eqref{eq:RetG} has the full time evolution from $\ql \qs \qr$ and the external environments).

Employing Eq.~\eqref{eq:essGr} in Eq.~\eqref{eq:CurrPC} (instead of using Eq.~\eqref{eq:GenCurr_PC}) gives exactly Eq.~\eqref{eq:Curr_SG_PC}. We stress that this applies to {\em many-body} or non-interacting systems within proportional coupling. We note also that the expression, Eq.~\eqref{eq:Curr_SG_PC}, is well-behaved in the continuum limit, since the coupling constants squared are related to canonical transforms and they decay with the inverse reservoir size. Also, we expect more general cases---beyond proportional coupling---of the large $\gamma$ regime can be analyzed, but the form will change since the single sum over $k\in\ql$ reflects the proportional coupling. Interestingly, a straightforward result, valid for arbitrary many-body or non-interacting systems, with or without proportional coupling, and with homogeneous or inhomogeneous relaxation results directly from an analysis on Eq.~\eqref{eq:Curr}~\footnote{This can be found by considering Eq.~\eqref{eq:essGr} and a similar equation for the lesser Green's function. Replacing the latter in Eq.~\eqref{eq:essGr} yields $\im 8 \pi n_i \delta_{ij}/\gamma_k$. Using both expressions in Eq.~\eqref{eq:Curr} gives $I = (I_\ql-I_\qr)/2$ with $I_\ql \approx 4 e \sum_{k\in\ql} \left( \tilde{f}_k \ik - \bkFoot \hat{n} \kkFoot \right)/\gamma_k$ and $I_\qr$ is identical in form (note that we do not assume proportional coupling, so the sum goes over different states, etc.)~and where $\hat{n}$ is a diagonal matrix of occupations for $\qs$, i.e., the diagonal of the correlation matrix $\qc$. This is a useful identity that can be used to validate many-body calculations, where the steady-state occupancies are extracted from the simulation and then the expression applied to ensure the right large-$\gamma$ behavior.}.

It is not a surprise that the large relaxation regime has only a minor dependence on the system---only through the coupling to the reservoir modes but not on the levels and interactions within the system itself. The relaxation overdamps the coherence that is forming between the reservoir modes and the system, giving an effective coupling that can be thought of perturbatively and hence the form $\ik / \gamma_k$. Strong local dephasing has a similar effect and yields a diffusion equation~\cite{dziarmaga_non-local_2012}. Coherence is needed for current to flow and thus the relaxation is limiting the flow from the reservoir states into the system. That being said, some alternative asymptotic regimes can appear for non-Markovian relaxation, see, e.g., Ref.~\onlinecite{gruss_communication_2017}. This has to do with the band structure of the reservoirs (e.g., band gaps), the level energies of the system, and symmetries.

\section{Conclusion}
\label{sec:Conc}

We are now in the era of ERAs: Extended reservoir approaches to transport are becoming widespread and employed in various applications, including to many-body transport~\cite{wojtowicz_open-system_2020,brenes_tensor-network_2020,fugger_nonequilibrium_2020,lotem_renormalized_2020}. We show that for local relaxation (i.e., acting independently on each of the modes in the reservoir) that a simple analytic expression results for both many-body and non-interacting systems with finite reservoirs proportionally coupled to the system. This will allow quite large finite reservoirs to be examined, especially for convergence to the Landauer (for non-interacting systems) or Meir-Wingreen (for many-body systems) limits~\cite{elenewski_performance_nodate}, as well as understanding underlying mechanistic behavior of finite representations of the reservoirs~\cite{wojtowicz_dual_nodate}. The asymptotic forms we derive also will be quite useful in benchmarking and validating numerical implementations, as well the full solutions, Eq.~\ref{eq:Curr} and Eq.~\ref{eq:nonintCurr}. We expect, therefore, that the results here will help support the use of extended, mesoscopic, or auxiliary reservoir simulations broadly---including its non-interacting incarnation as the DLvN---and bring it to bear on new applications.

\section*{Acknowledgments}

We thank Marek Rams, Gabriela W\'{o}jtowicz, Justin Elenewski, Yonatan Dubi, and Vladimir Aksyuk for helpful comments.

\vspace{0.2cm}

\section*{Data Availability}

Data sharing is not applicable to this article as no new data were created or analyzed in this
study.

\vspace{0.2cm}
\appendix 

\section{Equation-of-motion method}
\label{sec:EOM}
Due to the non-interacting nature of the reservoirs, the Dyson equations, Eqs.~\eqref{eq:Dysonjk} and~\eqref{eq:Dysonkj}, are easily obtained by the equation-of-motion method starting from the time-ordered Green's function:
\begin{align} 
\bG^T_{nm}(t-t^\p) = & -\im \avg{\textrm{T} \left\{ c_n (t) c_m^\dg (t^\p) \right\}} \label{eq:timeordered} \\
 = & -\im \Theta(t-t^\p) \avg{ c_n (t) c_m^\dg (t^\p)} \notag \\
 & +\im \Theta(t^\p-t) \avg{ c_m^\dg (t^\p) c_n (t)}  \notag 
\end{align}
with $\textrm{T}$ the time-ordering operator and $n,m \in \ql\qs\qr$. The equation of motion is 
\begin{align} 
-\im \partial_{t^\p} \bG^T_{nm}(t-t^\p) = & \delta(t-t^\p) \delta_{nm} \notag \\
 & - \im \Theta(t-t^\p) \avg{ c_n (t) \left[ H, c_m^\dg \right]_{t^\p} } \notag \\
 &+ \im \Theta(t^\p-t) \avg{ \left[ H, c_m^\dg \right]_{t^\p} c_n (t) } \label{eq:EOMto}, 
\end{align}
where we used $\avg{c_n (t) c_m^\dg (t^\p) + c_m^\dg (t^\p) c_n (t) }_{t \to t^\p} = \delta_{nm}$.  In Eq.~\eqref{eq:EOMto}, the operators should be evolved according to the full Hamiltonian (including also the {\em implicit} reservoirs). However, we can use that any single-particle extension of the reservoirs (e.g., their connection to the implicit reservoirs) requires just replacing the reservoir Green's functions with their dressed versions~\cite{haug_quantum_2008}, which is the approach we take here. 

For the current-carrying correlations, Eq.~\eqref{eq:EOMto} gives 
\begin{align} 
-\im \partial_{t^\p} \bG^T_{jk}(t-t^\p)  = & - \im \Theta(t-t^\p) \avg{ c_j (t) \left[ H, c_k^\dg \right]_{t^\p} } \notag \\
& + \im \Theta(t^\p-t) \avg{ \left[ H, c_k^\dg \right]_{t^\p} c_j (t) } \notag \\
= & \omega_k \bG^T_{jk} + \sum_{i\in\qs} v_{ik} \bG^T_{ji} 
\end{align}
with $j \in \qs$ and $k \in \ql \qr$. The solution is 
\[
\bG^T_{jk}(t-t^\p)= \sum_{i\in\qs} \int dt_1 \bG^T_{ji}(t-t_1) v_{ik} g_k^T (t_1-t^\p) .
\]
This expression has the same form as the contour-ordered Green's function~\cite{haug_quantum_2008}. The Keldysh rules give the lesser and retarded components~\cite{haug_quantum_2008}. After a Fourier transform, assuming a stationary state, this yields Eqs.~\eqref{eq:Dysonjk} or~\eqref{eq:Dysonkj} for the lesser component.

\end{document}